\documentclass[mathleft
]{an}
\usepackage{graphicx}
\usepackage{times}
\usepackage{natbib}
\bibpunct{(}{)}{;}{a}{}{}
\usepackage{url}
\newcommand{\aap}{A\&A}

\def\kms{$\mathrm{km\, s^{-1}}$}

\newcommand{\COBOLD}{{\sf CO$^5$BOLD}}
\newcommand{\cobold}{\COBOLD}

\newcommand{\xx}{\ensuremath{\mathrm{1D}_{\mathrm{LHD}}}}
\newcommand{\mD}{\ensuremath{\left\langle\mathrm{3D}\right\rangle}}
\newcommand{\loggf}{\ensuremath{\log\,gf}}
\DeclareRobustCommand{\ion}[2]{%
\relax\ifmmode
\ifx\testbx\f@series
{\mathbf{#1\,\mathsc{#2}}}\else
{\mathrm{#1\,\mathsc{#2}}}\fi
\else\textup{#1\,{\mdseries\textsc{#2}}}%
\fi}

\begin{document}

\Pagespan{789}{}
\Yearpublication{2010}%
\Yearsubmission{2010}%
\Month{11}%
\Volume{999}%
\Issue{88}%

\title{The solar photospheric abundance of zirconium} 

\author{E. Caffau\inst{1,2}\fnmsep\thanks{Corresponding author:
 Elisabetta Caffau - Gliese Fellow\newline \email{Elisabetta.Caffau@obspm.fr}\newline}
\and  R. Faraggiana\inst{3}
\and  H.-G. Ludwig\inst{1,2}
\and  P. Bonifacio\inst{2,4}
\and  M. Steffen \inst{5,2}
}
\titlerunning{Zirconium abundance in the Solar photosphere}
\authorrunning{E. Caffau el at.}
\institute{
Zentrum f\"ur Astronomie der Universit\"at Heidelberg, Landessternwarte, K\"onigstuhl 12, 69117 Heidelberg, Germany
\and 
GEPI, Observatoire de Paris, CNRS, Universit\'e Paris Diderot, Place
Jules Janssen, 92190
Meudon, France
\and 
Universit\`a degli Studi di Trieste,
via G.B. Tiepolo 11, 34143 Trieste, Italy
\and
Istituto Nazionale di Astrofisica,
Osservatorio Astronomico di Trieste,  Via Tiepolo 11,
I-34143 Trieste, Italy
\and
Astrophysikalisches Institut Potsdam, An der Sternwarte 16, D-14482 Potsdam, Germany
}

\received{30 May 2010}
\accepted{11 Nov 2010}
\publonline{later}

\keywords{Sun: abundances -- Stars: abundances -- Hydrodynamics -- Line: formation}

\abstract{%
Zirconium (Zr), together with strontium and yttrium, is an important element in the
understanding of the Galactic nucleosynthesis. In fact, the triad Sr-Y-Zr constitutes
the first peak of s-process elements. Despite its general relevance not many
studies of the solar abundance of Zr were conducted.
We derive the zirconium abundance in the solar photosphere
with the same CO5BOLD hydrodynamical model of the solar atmosphere
that we previously used to investigate the abundances of C-N-O. 
We review the zirconium lines available in the observed
solar spectra and select a sample of lines to determine the zirconium
abundance, considering lines of neutral and singly
ionised zirconium.  
We apply different line profile fitting strategies for a reliable
analysis of Zr lines that are blended by lines of other elements.
The abundance obtained from lines of neutral zirconium is
very uncertain because these lines are commonly blended and weak in the
solar spectrum.  However, we believe that some lines of ionised zirconium 
are reliable abundance indicators.
Restricting the set to \ion{Zr}{ii} lines, from the CO5BOLD
3D model atmosphere we derive A(Zr)=$2.62\pm 0.06$, where the quoted error is the RMS
line-to-line scatter.}
\maketitle

\section{Introduction}

\sloppy
The photospheric solar abundances of some elements have been studied
more often than others; this is mainly due to the importance of the
elements to explain nucleosynthesic processes. Moreover, the
difficulty of extracting suitable lines with accurately known
transition probabilities in the visible range of the solar spectrum
can also explain why some elements have been studied less extensively.

For the triad Sr-Y-Zr, there exist only a few detailed studies of their
solar abundances (an ADS\footnote{The Astrophysics Data System,
ads.harvard.edu} search yields 4 papers for Sr, 5 for Y, and 10 for
Zr) in spite of their importance in Galactic chemical evolution.
Sr-Y-Zr, Ba-La, and Pb are located at three abundance peaks of the
s-processes producing the enrichment of these elements in the Galaxy.
Knowledge of the present Sr-Y-Zr abundances in stars of different
metallicity and age is required to understand the complicated
nucleosynthesis of these elements (for details see
\citealt{travaglio04}).

Zirconium is present in the solar spectrum with lines of \ion{Zr}{i}
and \ion{Zr}{ii}.  The dominant species is \ion{Zr}{ii}. According to
our 3D model, and in agreement with our 1D reference model, about 99\%
of zirconium is singly ionised in the solar photosphere. 
Departure from local thermodynamical
equilibrium (LTE) probably affects \ion{Zr}{i} lines, producing a too
low Zr abundance under the assumption of LTE.  In fact NLTE effects,
even on weak unsaturated \ion{Zr}{i} lines, have been found by
\citet{brown83} 
for G and K giants.  However, both \citet{biemont81} and
\citet{bogdanovich96} analysed in LTE \ion{Zr}{i} lines (34 and 21
lines, respectively) and \ion{Zr}{ii} lines (24 and 15 lines,
respectively) in the solar photosphere, and found an excellent
agreement between the abundances derived from both ionisation stages.
Very recently \citet{velichko10} performed a NLTE analysis of 
zirconium in the case of the Sun and late type stars.
According to their computations the NLTE abundance is larger than the LTE one,
by up to 0.03\,dex for \ion{Zr}{ii}
and by 0.29\,dex for \ion{Zr}{i}.

Owing to the small number of \ion{Zr}{i} and \ion{Zr}{ii} lines in
metal poor stars \citep{gratton94}, 
it is important to analyse both of them in the solar photosphere to
derive the solar abundance from both ionisation stages and to assess
the agreement between the derived abundances.  In spectra of cool
stars it is easier to observe \ion{Zr}{i} lines.  For example,
\citet{goswami10} 
realised that none of the \ion{Zr}{ii} lines were usable in their
analysis of the cool Pop.\,II CH star HD~209621, and the Zr abundance
is derived from the only \ion{Zr}{i} line in their spectrum at
613.457\,nm.  A similar situation had been encountered by
\citet{vanture02} in their study of the Zr/Ti abundance ratio in cool S stars, where
only \ion{Zr}{i} lines in the red part of the spectrum can be used.

\section{Lines and atomic data}

According to \citet{malcheva06}, zirconium has five stable isotopes.
Four of them ($^{90}$Zr, $^{91}$Zr, $^{92}$Zr, and $^{94}$Zr) are
produced by the s-process; the fifth, $^{96}$Zr,
is produced in the r-process. Zirconium is a very refractory element, 
it is difficult to be vaporised by conventional
thermal means and, consequently, has been relatively little 
investigated in the laboratory.

In Table\,\ref{datazri} and Table\,\ref{datazrii} the
\ion{Zr}{i} and \ion{Zr}{ii} lines, with the \loggf\ values
available in the literature are collected.


\citet{biemont81} selected lines that lie at $\lambda$ $<$ 800\,nm.
They derived \loggf\ using the technique developed by
\citet{hannaford81}, suitable for highly refractory elements like Zr, to determine lifetimes
of 34 levels of \ion{Zr}{i} and 20 levels of \ion{Zr}{ii}. From these
measurements, coupled with the measurements of branching ratios, they
derived the \loggf\ values of 38 \ion{Zr}{i} and 31 \ion{Zr}{ii} lines.

\citet{bogdanovich96} computed and used \loggf\ values for 21 \ion{Zr}{i}
lines.  They give new \loggf\ values for 15 \ion{Zr}{ii} lines among
those selected by \citet{biemont81}.  Their \loggf\ are given in
Table\,\ref{datazrii}.

\citet{sikstrom99} derived the abundance of zirconium in HgMn star $\chi$ Lupi,
finding a disagreement when using \ion{Zr}{ii} or \ion{Zr}{iii} lines.
They measured the f-values for several \ion{Zr}{ii} lines in the UV at $\lambda < 300$\,nm.

\citet{vanture02} extended the search for \ion{Zr}{i} lines in the
near-IR, at wavelengths longer than 800\,nm, a region important in
the study of cool stars, because this is the region were cool stars
emit most of their flux.  The complete sample of these near-infrared
lines is also given in Table\,\ref{datazri}.  Because absorption bands
of molecules are weak in the warmer S stars, these near-IR zirconium
lines are better abundance indicators than the \ion{Zr}{ii} lines
in the blue part of the spectrum. These lines are, however, not
necessarily good for the Sun.

We have also checked the 13 \ion{Zr}{i} lines identified by
\citet{swensson70} in the \citet{delbouilleir} atlas, but 12 of them
appear severely blended. Only the line at 784.9\,nm is retained in
Table\,\ref{datazri}, as it was by \citet{biemont81} and
\citet{vanture02}.

In the present analysis we adopt the \loggf\ values of
\citet{biemont81} for all \ion{Zr}{i} lines.


According to the NIST database, \ion{Zr}{ii} lines lie only in the near UV-blue 
(241.941-535.035\,nm); 
only two weak lines (at 667.801 and 678.715\,nm) are present in NIST with 
$\lambda >$540\,nm.

The line identifications by \citet{moore66} on the Utrecht 
solar atlas have been
used by \citet{biemont81} to select 31 \ion{Zr}{ii} lines, 
seven of which were later discarded
because the derived abundance exceeded the mean by more than 3$\sigma$,
so they are likely blended with other unknown species. 
\citet{gratton94} used the same line list as \citet{biemont81}.
\citet{ljung06} derived new oscillator strengths 
for 263 \ion{Zr}{ii} lines and studied 7 
lines, the lines that they judged to be the best 
and unperturbed in the photospheric spectrum,  
to derive the solar Zr abundance based on both 
1D and 3D model atmospheres. 

We extracted from the 243 lines by \citet{malcheva06}, those in common
with \citet{biemont81}; for these lines the best values of \loggf,
according to \citet{malcheva06}, are the same as in \citet{biemont81}.
We compared the new \loggf\ values determined by \citet{ljung06} with
those by \citet{biemont81}. There is a good agreement for most of the
lines, with a few exceptions (see Table\,\ref{datazrii}).  We report
all these values in Table\,\ref{datazrii}.

In the present analysis we used the \loggf\ values of \citet{ljung06}
for all the \ion{Zr}{ii} lines.

\begin{table}
\caption{Lines of \ion{Zr}{i} chosen by \citet{biemont81} and \citet{vanture02}}
\label{datazri}
\begin{tabular}{lrrrrr}
\hline
\noalign{\smallskip}
 $\lambda$ &  E$_{\rm low}$ &  EW &  \multicolumn{3}{c}{\loggf} \\
 nm        &  eV           &  pm &  B & VW & Bog \\
\noalign{\smallskip}
\hline
\noalign{\smallskip}
 350.9331  &  0.07 &  0.65  & --0.11 &          & --0.21 \\ 
 360.1198  &  0.15 &  1.3   & --0.47 &          &        \\
 389.1383  &  0.15 &  1.7   & --0.10 &          &        \\
 402.893   &  0.52 &  0.06  & --0.72 &          &        \\
 403.0049  &  0.60 &  0.26  & --0.36 &          & --0.59 \\
 404.3609  &  0.52 &  0.58  & --0.37 &          &        \\
 407.2696  &  0.69 &  0.57  &  +0.31 &          &  +0.24 \\
 424.1706  &  0.65 &  0.37  &  +0.14 &          &  +0.07 \\
 450.7100  &  0.54 &  0.36  & --0.43 &          & --0.46 \\
 454.2234  &  0.63 &  0.46  & --0.31 &          &        \\
 468.7805  &  0.73 &  1.00  &  +0.55 &          &  +0.30 \\
 471.0077  &  0.69 &  1.05  &  +0.37 &          &  +0.19 \\
 473.2323  &  0.63 &  0.25  & --0.49 &          & --0.56 \\
 473.9454  &  0.65 &  0.55  &  +0.23 &          &  +0.07 \\
 477.2310  &  0.62 &  0.53  &  +0.04 &          & --0.07 \\
 478.494   &  0.69 &  0.16  & --0.49 &          &        \\
 480.589   &  0.69 &  0.15  & --0.42 &          & --0.63 \\
 480.9477  &  1.58 &  0.16  &  +0.16 &          &        \\
 481.5056  &  0.65 &  0.20  & --0.53 &          & --0.22 \\
 481.5637  &  0.60 &  0.30  & --0.03 &          &        \\
 482.806   &  0.62 &  0.19  & --0.64 &          &        \\
 504.655   &  1.53 &  0.050 &  +0.06 &          & --0.25 \\
 538.5128  &  0.52 &  0.18  & --0.71 &          &        \\
 612.746   &  0.15 &  0.21  & --1.06 &  --1.06  & --0.87 \\
 613.457   &  0.00 &  0.19  & --1.28 &  --1.28  & --1.05 \\
 614.046   &  0.52 &  0.073 & --1.41 &  --1.41  & --0.85 \\
 614.3183  &  0.07 &  0.21  & --1.10 &  --1.10  & --0.98 \\
 631.303   &  1.58 &  0.11  &  +0.27 &          &  +0.18 \\
 644.572   &  1.00 &  0.094 & --0.83 &  --0.83  &        \\
 699.084   &  0.62 &  0.050 & --1.22 &          & --1.44 \\
 709.776   &  0.69 &  0.21  & --0.57 &          &        \\
 710.289   &  0.65 &  0.065 & --0.84 &          & --1.06 \\
 743.989   &  0.54 &        &        &  --1.18  &        \\
 755.149   &  1.58 &        &        &  --1.36  &        \\
 755.473   &  0.51 &        &        &  --2.28  &        \\
 755.841   &  1.54 &        &        &  --1.47  &        \\
 756.213   &  0.62 &        &        &  --2.71  &        \\
 781.935   &  1.82 &  0.065 & --0.38 &  --0.39  &        \\
 782.292   &  1.75 &        &        &  --1.14  &        \\
 784.938   &  0.69 &  0.10  & --1.30 &  --1.30  &        \\
 856.859   &  0.73 &        &        &  --2.80  &        \\
 857.1085  &  1.53 &        &        &  --2.07  &        \\
 858.421   &  1.86 &        &        &  --1.32  &        \\
 858.787   &  1.48 &        &        &  --2.12  &        \\
 874.958   &  0.60 &        &        &  --2.79  &        \\
\noalign{\smallskip}
\hline
\noalign{\smallskip}
 350.1133  &  0.07 &  0.7   & --0.93 &\\
 357.5765  &  0.07 &  3.3   & --0.03 &\\
 366.3698  &  0.15 &  2.5   &  +0.01 &\\
 588.5629  &  0.07 &  0.050 & --2.12 &          & -1.82 \\
\noalign{\smallskip}
\hline
\noalign{\smallskip}
\end{tabular}
\\
B: \citet{biemont81}\\
VW: \citet{vanture02}\\
Bog: \citet{bogdanovich96}\\
\end{table}

\begin{table}
\caption{Lines of \ion{Zr}{ii} chosen by \citet{biemont81} and \citet{ljung06}}
\label{datazrii}
\begin{tabular}{lrrrrrrr}
\hline
\noalign{\smallskip}
 $\lambda$ &  E$_{\rm low}$ &  \multicolumn{2}{c}{EW [pm]} &  \multicolumn{3}{c}{\loggf} \\
 nm        &  eV           &  B & L &  B & Bog &  L \\
\noalign{\smallskip}
\hline
\noalign{\smallskip}
 343.2415 & 0.93  &  2.1  &       &  --0.75 &  --0.51 & --0.72 \\
 345.4572 & 0.93  &  1.00 &       &  --1.34 &         & --1.33 \\
 345.8940 & 0.96  &  1.6  &       &  --0.52 &         & --0.48 \\
 347.9017 & 0.53  &  2.8  &       &  --0.69 &  --1.12 & --0.67 \\
 347.9393 & 0.71  &  5.1  &       &   +0.17 &   +0.12 &  +0.18 \\
 349.9571 & 0.41  &  2.4  &       &  --0.81 &  --1.08 & --1.06 \\
 350.5666 & 0.16  &  5.1  &       &  --0.36 &  --0.62 & --0.39 \\
 354.9508 & 1.24  &  1.6  &       &  --0.40 &  --0.68 & --0.72 \\
 355.1951 & 0.09  &  5.8  &       &  --0.31 &         & --0.36 \\
 358.8325 & 0.41  &  2.7  &       &  --1.13 &  --1.25 & --1.13 \\
 360.7369 & 1.24  &  1.3  &       &  --0.64 &  --0.48 & --0.70 \\
 367.1264 & 0.71  &  3.2  &       &  --0.60 &  --0.56 & --0.58 \\
 371.4777 & 0.53  &  3.0  &       &  --0.93 &         & --0.96 \\
 379.6496 & 1.01  &  1.5  &       &  --0.83 &  --1.17 & --0.89 \\
 383.6769 & 0.56  &  4.7  &       &  --0.06 &  --0.22 & --0.12 \\
 403.4091 & 0.80  &  0.65 &       &  --1.55 &         & --1.51 \\
 405.0320 & 0.71  &  2.38 & 2.20  &  --1.00 &  --0.60 & --1.06 \\ 
 408.5719 & 0.93  &  0.54 &       &  --1.61 &  --1.54 & --1.84 \\
 420.8980 & 0.71  &  4.3  & 4.26  &  --0.46 &         & --0.51 \\ 
 425.8041 & 0.56  &  2.6  & 2.34  &  --1.13 &         & --1.20 \\ 
 431.7321 & 0.71  &  1.20 &       &  --1.38 &         & --1.45 \\
 444.2992 & 1.49  &  2.0  & 2.04  &  --0.33 &         & --0.42 \\ 
 449.6962 & 0.71  &  3.6  & 3.15  &  --0.81 &  --0.87 & --0.89 \\ 
 511.2270 & 1.66  &  0.83 & 0.78  &  --0.59 &  --0.58 & --0.85 \\ 
\noalign{\smallskip}
\hline
\noalign{\smallskip}
 363.0027 & 0.36  &  3.8  &       &  --1.11 &         & --1.11 \\
 407.1093 & 1.00  &  2.1  &       &  --1.60 &         & --1.66 \\
 414.9202 & 0.80  &  7.5  &       &  --0.03 &         & --0.04 \\
 416.1208 & 0.71  &  5.8  &       &  --0.72 &         & --0.59 \\
 426.4925 & 1.66  &  1.50 &       &  --1.41 &         & --1.63 \\
 444.5849 & 1.66  &  0.83 &       &  --1.35 &         &        \\
 461.3921 & 0.97  &  2.91 &       &  --1.52 &         & --1.54 \\
\noalign{\smallskip}
\hline
\noalign{\smallskip}
 402.4435 & 0.999 &       &  1.20 &         &         & --1.13 \\
\noalign{\smallskip}
\hline
\noalign{\smallskip}
\end{tabular}
\\
B: \citet{biemont81}\\
Bog: \citet{bogdanovich96}\\
L:  \citet{ljung06} 
\end{table}

\section{Solar Zr abundance in the literature}

Several analyses of the solar  Zr abundance,
made before 1980, used \citet{corliss62} \loggf\ 
(\citealt{aller65}, 
\citealt{wallerstein66}, \citealt{grevesse68})  
or  older ones \citep{goldberg60} 
or have been made to derive 
a temperature correction for the \citet{corliss62} data \citep{allen76}.  
The use of these transition probabilities, which are known to
contain errors depending on excitation and temperature, coupled with the use
of old solar spectral atlases, explains discrepancies of these Zr abundance
determinations based either on \ion{Zr}{i} and/or \ion{Zr}{ii}.

We concentrate on the more recent determinations
of the photospheric solar Zr abundance, 
that we summarise below.

\begin{enumerate}
\item
\citet{biemont81} analysed the \citet{delbouille} atlas.  The
equivalent widths (EWs) were independently measured by E.\,Bi\`emont
and N.\,Grevesse, and the results examined and discussed to produce
the final list.  They used the Holweger-M\"uller solar model
\citep{hhsunmod,hmsunmod} with a microturbulence, $\xi$, of $0.8$\,\kms.
They note that their analysis is independent of line-broadening
parameters since most of the lines are faint or very faint.  The
\loggf\ are from their own measurements.  They find that the
abundance from \ion{Zr}{i} lines is independent of $\xi$ and that
the abundance from \ion{Zr}{ii} lines is insensitive to the model.
These authors rejected 4 \ion{Zr}{i} lines and 7 \ion{Zr}{ii} lines
from their selected sample in their final solar abundance analysis
because they imply a Zr abundance which is more than 3$\sigma$
higher than the average. These lines appear after the horizontal
line in Tables\,\ref{datazri} and \ref{datazrii}.  They conclude that
\ion{Zr}{ii} lines are better abundance indicators. They obtain 
a zirconium abundance of A(Zr)=$2.57\pm 0.07$ from \ion{Zr}{i} 
(34 lines) and A(Zr)=$2.56\pm 0.07$ from \ion{Zr}{ii} (24 lines).

\item 
\citet{gratton94} studied the Zr behaviour in metal-poor stars.
They used as model atmosphere a grid from \citet{bell76} for giants,
and a similar grid for dwarfs provided by Bell.  They also analysed
the solar spectrum, to have a reference A(Zr)$\odot$.  For this
purpose they used the same 24 lines as \citet{biemont81}, the
Holweger-M\"uller model with microturbulence of $1.5$\,\kms, and the solar
abundances from \citet{anders89}.  They remark that in their metal
poor stars Zr abundances from \ion{Zr}{i} are, on average, lower than
those from \ion{Zr}{ii} lines, a result similar to that found by
\citet{brown83} for Pop.\,I stars.  For the analysis of the metal poor
stars they used 7 \ion{Zr}{ii} lines and 3 of them (407.1, 414.9, and
416.1\,nm) are among those discarded by \citet{biemont81}, suggesting
that these are no longer significantly blended in metal poor stars.
They obtain for the solar Zr abundance 2.59$\pm$0.04 from \ion{Zr}{i}
(34 lines) $\sigma$=0.22 dex, 2.53$\pm$0.03 from \ion{Zr}{ii} (24
lines) $\sigma$=0.14 dex, respectively.

\item 
\citet{bogdanovich96} computed theoretical \loggf, and 
used the Holweger-M\"uller model with $\xi$=0.8, and a
model computed with a code by \citet{kipper81}
which is a modified version of the ATLAS 5 code \citep{Kurucz70}. 
They used the  damping constants modified by \citet{galdikas88}, 
and the EW measurements of \citet{biemont81}.
The results are: A(Zr)=2.60$\pm$0.07 from \ion{Zr}{i} (21 lines) and
A(Zr)=2.61$\pm$0.11 from \ion{Zr}{ii} (15 lines).
They adopted A(Zr)$=2.60\pm 0.06$.

\item
\citet{ljung06} considered only \ion{Zr}{ii} lines, they used
\loggf\ from experimental branching ratios measured by them, and
radiative lifetimes taken from different papers.  They used 1D
Holweger-M\"uller and MARCS models, and a 3D model computed with the
Stein-Nordlund code \citep{stein} for line profile fitting.  EWs of
the best 3D fitting profiles are used to estimate the abundance.
From 7 \ion{Zr}{ii} lines, six of which in common with
\citet{biemont81}, they obtain A(Zr)=$2.58\pm 0.02$ with the 3D model,
A(Zr)=$2.56\pm 0.02$ with MARCS model, and A(Zr)=$2.63\pm 0.02$ with
the HM model. Their adopted value is A(Zr)=$2.58\pm 0.02$.

\item
\citet{velichko10} performed the first NLTE analysis of Zr in the solar photosphere.
They investigated all the Zr lines analysed by \citet{biemont81}
and \citet{ljung06} and selected a subsample of two \ion{Zr}{i}
lines (424.1 and 468.7\,nm) and ten \ion{Zr}{ii} lines 
(347.9, 350.5, 355.1, 405.0, 420.8, 425.8, 444.2, 449.6, and 511.2\,nm).
Analysing the \citet{kuruczflux} solar spectrum
with a MAFAGS model atmosphere (5780\,K/4.44/0.0) and a microturbulence of 0.9\kms
they derive a LTE abundance of 2.33 and 2.61 from \ion{Zr}{i} and \ion{Zr}{ii}
lines, respectively.
They constructed a Zr model atom and derived the NLTE abundance
with different assumptions about the rates of collisions with H-atoms.
In their scenario, the NLTE abundance is always higher than the LTE abundance
varying in a range from 0.21 to 0.32\,dex for \ion{Zr}{i} and 
from 0.01 to 0.08\,dex for \ion{Zr}{ii}, depending on the adopted cross sections for 
collision with H-atoms.
Their best estimate of the NLTE abundance (correction) 
from \ion{Zr}{i} and \ion{Zr}{ii} lines is 2.62 (0.29) and 2.64 (0.03) dex, respectively.
Their adopted Zr abundance is $2.63\pm 0.07$.
\end{enumerate}

The abundance of Zr in the solar system is derived from the analysis
of meteorites. Zirconium is not a volatile element, and its abundance
in meteorites should agree with the one in the solar photosphere. 
To compare the abundances derived from the meteorites, 
on the cosmochemical abundance scale relative to $10^6$ Si atoms,
with the ones derived form the solar photosphere, 
on the astronomical scale of $10^{12}$ H-atoms,
a coupling factor must be derived.
\citet{lodders03} and \citet{lodders09} to link the two scales introduce an average factor
by looking at a sample of refractory elements well-determined in the photosphere
and at the same elements in meteorites.
In this way the factor is not affected by the individual uncertainty
and, if the abundance of one element derived from the photosphere changes, 
the conversion factor is not much affected.

The meteoritic Zr abundance is A(Zr)=$2.57\pm 0.04$, \citet{lodders09}. 
Other Zr meteoritic values that can be found in the literature are:
$2.61\pm 0.03$ according to \citet{anders89},
$2.61\pm 0.02$ according to \citet{grevesse98},
$2.60\pm 0.02$ according to \citet{lodders03}.
The value of $2.53\pm 0.04$ given in \citet{asplund09} is based on the same
data from \citet{lodders09}, but they give a lower value for Zr
on the astronomical scale because they use a
lower coupling factor for the cosmochemical and astronomical scales.
In the same way the value of $2.57\pm 0.02$ in \citet{asplund05} and \citet{grevesse07}
is based on data from \citet{lodders03}, but their value is lower because 
again a different scaling factor (which is 0.03 $\log$ units smaller than
the one recommended in \citet{lodders03}) was used.

\section{Model atmospheres}
We base our abundance analysis on the same model atmospheres that
we used in the previous analysis of solar abundance determinations,
summarised in  \citet{solarphy}.
We rely on a time-dependent, 3D hydrodynamical model atmosphere,
computed with the \cobold\ code; see \citet{freytag02}  
and \citet{freytag03} for details.
This model has a box size of $5.6\times 5.6\times 2.27\,{\rm Mm}^3$,
resolved by $140\times 140\times 150$ grid points.
Its range in Rosseland optical depth is $-6.7<\log\tau_{\rm Ross}<5.5$.
The complete time series is formed by 
90 snapshots covering 1.2\,h of solar time.
For the spectral synthesis computations we selected 19 representative snapshots
out of the complete series of 90 snapshots.

To compute 3D-corrections for the 
zirconium abundance, we make use of two 1D reference atmospheres:
the 1D model obtained by horizontally averaging each 3D snapshot over
surfaces of equal (Rosseland) optical depth (henceforth \mD\ model), 
and  the hydrostatic 1D mixing-length model computed with the LHD code, 
that employs the same micro-physics and radiative transfer scheme as 
the \cobold\ code, (henceforth \xx\ model).
We define 3D-corrections as in \cite{solarphy}, 
${\Delta_{\rm gran}={\rm A(Zr)}_{\rm 3D}}-{\rm A(Zr)}_{\rm \mD}$, 
to isolate the effects of horizontal fluctuations (granulation effects), 
and ${\Delta_{\rm LHD}={\rm A(Zr)}_{\rm 3D}}-{\rm A(Zr)}_{\rm LHD}$,
to measure the total 3D effect. 
For details see \cite{solarphy} and \cite{zolfito}.

We also considered the semi-empirical 
Holweger-M\"uller solar model \citep[][hereafter HM]{hhsunmod,hmsunmod}
for comparison.

\section{Observed spectra}

We analysed the same four solar spectral atlases, publicly available 
(two disc-centre and two disc-integrated atlases), that we used in all 
our previous works on solar abundances, such as in \citet{solarphy}.
These high resolution, high signal-to-noise ratio (S/N) spectra, are:
\begin{enumerate}
\item
the disc-integrated spectrum of \citet{kuruczflux}, based on fifty solar FTS
scans, observed by J. Brault and L. Testerman at Kitt Peak between
1981 and 1984;
\item
the FTS atlas of \citet{neckelobs}, observed at Kitt Peak in the 1980ies,
providing both disc-centre and disc-integrated data;
\item
the disc-centre atlas of \citet{delbouille}, observed from the Jungfraujoch.
\end{enumerate}

\section{Abundance determination methods}
\label{s:methods}
In principle, the best way to derive the abundance of an element from 
a spectral line is by line
profile fitting. Once the atomic data are known, this technique allows 
to take into account the strength of the line and its shape at the same time.
The limiting factor is that the observed solar spectra are of such a good
quality that the synthetic line profile is often not realistic enough 
to reproduce the shape of the line in the observed spectrum.
Synthetic line profiles derived from 1D model atmospheres do not take
into account the line asymmetry induced by convection, they are symmetric.
In many cases NLTE effects modify the line profile \citep[see][]{asplund04}.
In case the line is not clean, but some contaminating lines are
present in the range, the line profile fitting should take into account these 
blending components. Good atomic data for these blending lines are then 
necessary, and, as for the line of interest, granulation and NLTE effects 
can be limiting factors.

To fit a line blended with other components, one has to optimise the agreement
between synthetic and observed spectra also for the blending lines, allowing, 
in the line profile fitting process, the abundance of the blending lines to 
change as well. Even though the 3D spectrum synthesis is still computationally
demanding,
we have attempted to derive A(Zr) by properly taking
into account also the blending lines whenever possible.
In the fitting procedure, the Zr abundance and the abundances of the 
blending lines are adjusted independently until the best match of the 
observed spectrum is achieved (in the following referred to as method `A').

We also tested a simplified fitting approach (in the following method
'B'): we synthesised one typical 3D profile of each blending
components, in addition to a grid with different abundances for the Zr
line of interest. Then, in the fitting procedure, a profile is
interpolated in the grid for Zr, while the profiles (more precisely
the relative line depressions) of the blending lines are simply
scaled, and then the blending lines and the Zr line are added linearly
to obtain the composite line profile.  In the fitting procedure, the
Zr abundance and the scaling factors (and shifts) of the blending
lines are adjusted until the best agreement with the observed spectrum
is found.  An advantage of this fitting procedure is that it works
even if the atomic data of the blending components are poorly known.
Note, however, that this procedure is theoretically justified only if
all the lines are on the linear part of the curve of growth. If the
lines are partly saturated, the line strengths of all components are
underestimated, and so is the derived Zr abundance.  Nevertheless,
method 'B' was also applied to partly saturated lines in order to
understand its limitations.

Methods 'A' and 'B' provide the Zr abundance without the need to measure
an equivalent width. However, we can formally derive the equivalent width
of the Zr component from the synthetic line profile computed for Zr only
(ignoring all blends) with the Zr abundance that provides the best fit of
the observed spectrum. For method 'B', this is simply the equivalent width 
of the Zr line that gives the best agreement between
added-synthetic and observed profile.

A third alternative (method 'C') is the measurement of the equivalent
width of the line of interest, which is then translated into an abundance
via the 3D curve-of-growth. For this purpose, we employ the IRAF task
{\tt splot}\footnote{\url{http://iraf.noao.edu/}} for fitting Gaussian or
Voigt profiles to the observed line.
Even in the case that the line of interest is contaminated by 
unidentified components, which cannot be taken into account with methods 
'A' and 'B', it may still be possible to use {\tt splot} with the 
deblending option to determine the equivalent widths of all components. 
This procedure may also be useful if the number of blending lines is 
too large to be treated conveniently by the methods 'A' or 'B' based 
on synthetic spectra. However, as method 'B', this approach is only 
justified if all involved lines are weak.

\section{Zirconium abundance from \ion{Zr}{i} lines}

We looked at the \ion{Zr}{i} lines in the sample of \citet{biemont81}.
All these lines are rather weak and blended, or in a crowded spectral region.
After inspection, we analysed a subsample of 11 lines,
discarding the lines we think to be too heavily blended
and/or in a too crowded range.
We found our EW measurements to be affected by rather large uncertainties, 
but for the majority of the 11 lines we agree with the results of 
\citet{biemont81}.
We think that the cleanest lines are the ones at 
480.9, 481.50, 614.0, and 644.5\,nm, but also these lines are too poor
to give precise information on the solar abundance.
We found it difficult to use the flux spectra to measure the EWs of so 
weak and blended lines, and decided to use only the
disc-centre spectra, which are sharper.
From the four clean lines mentioned above, we obtain, A(Zr)$=2.65\pm 0.07$.
If we remove also the 644.5\,pm line we obtain
A(Zr)$=2.62\pm 0.04$. This result is in perfect agreement with the abundance
obtained from \ion{Zr}{ii} lines (see below), but we consider this agreement 
fortuitous. From the \xx\ model we find a slightly higher value of 
A(Zr)=$2.67\pm 0.07$, and from the \mD\ model A(Zr)=$2.74\pm 0.08$.

\section{Investigation of \ion{Zr}{ii} lines}

We picked up all the \ion{Zr}{ii} 
lines of \citet{biemont81} and the line at 402.4\,pm
analysed in \citet{ljung06}. 
Our starting sample was formed by 32 \ion{Zr}{ii} lines.
After inspection of the observed data, and comparison with 1D synthetic 
spectra, we discarded 11 lines that we think are too heavily blended.
After analysis, we discarded four other 
lines that also \citet{biemont81} rejected
because they provided a too high abundance.
Our final sample is formed 
by the seven lines of \citet{ljung06}, that we discuss
one by one in the next section, and 
10 lines that are used for the abundance analysis
in \citet{biemont81}.

\begin{figure}
\includegraphics[width=80mm,clip=true]{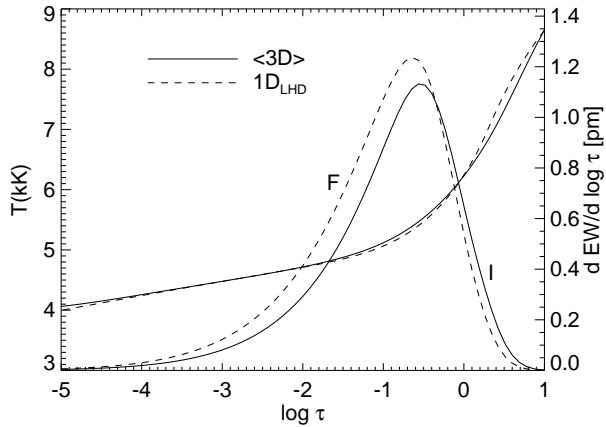}
\caption{Temporal and horizontal average of the temperature profile 
  of the 3D model (solid line) and the temperature profile of \xx\ model,
  shown as a function of (Rosseland) optical depth.
  In addition, the equivalent width contribution functions (lower, roughly 
  Gaussian-shaped curves) for the 405.0\,nm \ion{Zr}{ii} line are 
  plotted on the same optical depth scale
  for disc-centre (solid line) and disc-integrated spectra (dashed line).
}
\label{contf}
\end{figure}

These \ion{Zr}{ii} lines have lower level energies below 1.7\,eV.
If we look at the equivalent width 
contribution functions for disc-centre of the 3D model,
we can see that they are formed in a range in $\tau_{\rm Ross}$ between 
$-4$ and $0$, with the
maximum in the range $-1.0<\tau_{\rm Ross} < -0.5$ (see Fig.\,\ref{contf}).  
Only two lines (350.5 and 355.1\,nm) are formed over a larger range, with
a maximum around $\tau_{\rm Ross} \approx -2.0$, because these two lines 
have smaller lower level energies, and are significantly stronger than 
the others.

The abundance results we obtained from 
the \ion{Zr}{ii} lines are summarised in Table\,\ref{zr9abbo}.
We averaged the EW measured in the two disc-centre and the two 
disc-integrated spectra, respectively. 
The last column of the table indicates if the result refers to
disc-centre or disc-integrated spectra.
From the complete sample of 17 lines we obtain
A(Zr)=2.612$\pm$0.093, with $2.638\pm 0.078$ and $2.586\pm 0.100$
from disc-centre and disc-integrated spectra, respectively.
The line at 347.9\,nm gives a very low abundance, 
while the line at 355.1\,nm gives a rather high one.
Both these two lines lie in a very crowded 
region of the spectrum and it is very difficult
to place the continuum.
When we remove these two lines from 
the sample, we have 15 \ion{Zr}{ii} lines and we obtain
A(Zr)=2.616$\pm$0.060, with $2.643\pm 0.050$ and $2.589\pm 0.057$
from disc-centre and disc-integrated spectra, respectively.
We note that the 3D abundance from flux is systematically lower 
than the one from the intensity spectra.

When we look only at the seven lines from 
\citet{ljung06}, we find that the line-to-line
scatter is reduced with respect to the sample of 15 lines.
The 3D abundance we find is $2.653\pm 0.022$ and $2.623\pm 0.031$ for
disc-centre and disc-integrated data, respectively.
The abundance averaged over both disc-centre 
and disc-integrated spectra is $2.638\pm 0.031$.
We find a very low line-to-line scatter, but again 
the disc-centre spectra give a systematically higher abundance 
than the disc-integrated spectra, by an amount comparable to the 
line-to-line scatter of $\approx 0.03$\,dex.

We found similar results for Fe, Hf, and Th, where the
abundances derived from  disc-integrated spectra are
lower by 0.02, 0.05, and 0.02\,dex, respectively, than the
disc-centre abundances (see \citealt{thhf} and \citealt{solarphy} 
for details). 
A similar behaviour is seen when using 1D model atmospheres.
We have presently no obvious explanation for this result.
 
The zirconium abundance derived from 1D models depends on the choice 
of the microturbulence parameter $\xi$.
The microturbulence we find for the sample of 15 \ion{Zr}{ii} 
lines, by requiring that the relation between EW and A(Zr)
obtained with the 1D model has the same slope as that of the 3D model
(method 3a in \citet{mst09}), is consistent with the results obtained
by \citet{mst09} from a sample of \ion{Fe}{ii} lines.
We adopt $\xi = 1.0$\,\kms\ for the analysis of disc-integrated spectra,
and obtain A(Zr)$_{\mD}=2.597\pm 0.053$ and 
A(Zr)$_{\rm LHD}=2.575\pm 0.057$.
For disc-centre data, we choose $\xi = 0.7$\,\kms\ and
the result is A(Zr)$_{\mD}=2.652\pm 0.044$ and 
A(Zr)$_{\rm LHD}=2.635\pm 0.045$ (for $\xi=1.0$\,\kms\ 
we get A(Zr)$_{\mD}=2.602\pm 0.067$, A(Zr)$_{\rm LHD}=2.587\pm 0.072$).

3D corrections are small in absolute value and depend on the value of
the microturbulence. The average 3D correction with respect to the
reference \xx\ model is positive, $\Delta_{\rm LHD}=+0.011\pm 0.012$\,dex, 
while, with respect to the \mD\ model it is slightly negative, 
$\Delta_{\rm gran}=-0.008\pm 0.016$\,dex, again
taking $\xi=0.7$\,\kms\ for disc-centre and $\xi=1.0$\,\kms\ for
disc-integrated spectra.

With the HM model we obtain A(Zr)$_{\rm HM}=2.679\pm 0.044$
and A(Zr)$_{\rm HM}=2.619\pm 0.053$ for disc-centre and integrated disc, respectively.
Taking into account both lines of disc-centre and integrated disc, we obtain
A(Zr)$_{\rm HM}=2.649\pm 0.057$, and when applying the $\Delta_{\rm gran}$
correction, we have A(Zr)$_{\rm HM}=2.641\pm 0.057$.

For our final determination of the solar zirconium abundance,
we adopt the complete sample of 15 \ion{Zr}{ii} lines, and our 
recommended value is A(Zr)$_{\rm 3D}=2.62\pm 0.06$, obtained with 3D model atmosphere.

\begin{table*}
\caption{The results from the 7 lines of \ion{Zr}{ii} 
from \citet{ljung06}, and other 10 lines from \citet{biemont81}.}
\label{zr9abbo}
\begin{tabular}{ccrcccccrcc}
\hline
\noalign{\smallskip}
 $\lambda$ & E$_{\rm low}$ & EW & \loggf\ &  \multicolumn{4}{c}{A(Zr)} & \multicolumn{2}{c}{3D-Corrections} & Sp\\
     nm    &   eV          & pm &         & 3D & \mD\ & \xx\ & HM & $\Delta_{\rm gran}$ & $\Delta_{\rm LHD}$\\
           &               &    &         &    & \multicolumn{3}{c}{$\xi = 1.0/0.7$\,\kms}& \multicolumn{2}{c}{$\xi = 1.0/0.7$\,\kms}\\
\noalign{\smallskip}
\hline
\noalign{\smallskip}
    402.4453 &   0.999 &   1.33 &   --1.13 &    2.667 &    2.645/2.660 &    2.640/2.655 &    2.673/2.689 &     0.022/  0.007 &    0.027/  0.012 &  I  \\
    402.4453 &   0.999 &   1.43 &   --1.13 &    2.613 &    2.616/2.636 &    2.602/2.622 &    2.638/2.659 &   --0.003/--0.023 &    0.011/--0.009 &  F  \\
    405.0320 &   0.713 &   2.17 &   --1.06 &    2.617 &    2.588/2.620 &    2.577/2.608 &    2.617/2.649 &     0.029/--0.003 &    0.040/  0.009 &  I  \\
    405.0320 &   0.713 &   2.40 &   --1.06 &    2.583 &    2.593/2.637 &    2.573/2.615 &    2.617/2.662 &   --0.010/--0.054 &    0.010/--0.032 &  F  \\
    420.8980 &   0.713 &   4.22 &   --0.51 &    2.630 &    2.559/2.671 &    2.533/2.640 &    2.592/2.707 &     0.071/--0.040 &    0.098/--0.010 &  I  \\
    420.8980 &   0.713 &   4.45 &   --0.51 &    2.578 &    2.596/2.733 &    2.561/2.693 &    2.627/2.767 &   --0.018/--0.155 &    0.017/--0.116 &  F  \\
    425.8041 &   0.559 &   2.36 &   --1.20 &    2.642 &    2.613/2.648 &    2.597/2.631 &    2.641/2.677 &     0.029/--0.006 &    0.045/  0.011 &  I  \\
    425.8041 &   0.559 &   2.67 &   --1.20 &    2.627 &    2.640/2.691 &    2.616/2.665 &    2.664/2.716 &   --0.013/--0.064 &    0.011/--0.037 &  F  \\
    444.2992 &   1.486 &   2.19 &   --0.42 &    2.672 &    2.637/2.668 &    2.625/2.655 &    2.665/2.697 &     0.035/  0.004 &    0.047/  0.017 &  I  \\
    444.2992 &   1.486 &   2.38 &   --0.42 &    2.651 &    2.651/2.694 &    2.631/2.671 &    2.673/2.717 &   --0.000/--0.043 &    0.020/--0.020 &  F  \\
    449.6962 &   0.713 &   3.11 &   --0.89 &    2.661 &    2.618/2.674 &    2.596/2.649 &    2.647/2.704 &     0.042/--0.014 &    0.065/  0.012 &  I  \\
    449.6962 &   0.713 &   3.41 &   --0.89 &    2.642 &    2.655/2.733 &    2.626/2.699 &    2.680/2.761 &   --0.013/--0.091 &    0.016/--0.057 &  F  \\
    511.2270 &   1.665 &   0.84 &   --0.85 &    2.685 &    2.664/2.672 &    2.653/2.661 &    2.688/2.697 &     0.021/  0.013 &    0.032/  0.024 &  I  \\
    511.2270 &   1.665 &   0.92 &   --0.85 &    2.666 &    2.662/2.673 &    2.647/2.658 &    2.679/2.691 &     0.004/--0.007 &    0.019/  0.008 &  F  \\
\noalign{\smallskip}                                                                                                                         
\hline                                                                                                                                       
\noalign{\smallskip}                                                                                                                         
    345.4572 &   0.931 &   0.96 &   --1.33 &    2.781 &    2.760/2.772 &    2.748/2.760 &    2.779/2.791 &     0.021/  0.009 &    0.032/  0.021 &  I  \\
    345.4572 &   0.931 &   1.00 &   --1.33 &    2.687 &    2.687/2.702 &    2.672/2.686 &    2.702/2.717 &   --0.001/--0.016 &    0.015/  0.000 &  F  \\
    347.9017 &   0.527 &   2.80 &   --0.67 &    2.422 &    2.373/2.437 &    2.350/2.411 &    2.393/2.458 &     0.049/--0.015 &    0.072/  0.011 &  I  \\
    347.9017 &   0.527 &   2.90 &   --0.67 &    2.315 &    2.330/2.406 &    2.301/2.375 &    2.348/2.424 &   --0.015/--0.091 &    0.014/--0.059 &  F  \\
    349.9571 &   0.409 &   2.50 &   --1.06 &    2.598 &    2.559/2.609 &    2.536/2.585 &    2.579/2.630 &     0.040/--0.011 &    0.062/  0.014 &  I  \\
    349.9571 &   0.409 &   2.70 &   --1.06 &    2.525 &    2.540/2.606 &    2.513/2.576 &    2.558/2.624 &   --0.015/--0.081 &    0.012/--0.051 &  F  \\
    350.5666 &   0.164 &   5.10 &   --0.39 &    2.614 &    2.469/2.676 &    2.429/2.633 &    2.496/2.703 &     0.145/--0.062 &    0.185/--0.019 &  I  \\
    350.5666 &   0.164 &   5.40 &   --0.39 &    2.514 &    2.527/2.745 &    2.482/2.697 &    2.553/2.771 &   --0.013/--0.231 &    0.032/--0.183 &  F  \\
    354.9508 &   1.236 &   1.40 &   --0.72 &    2.656 &    2.627/2.646 &    2.613/2.632 &    2.645/2.665 &     0.029/  0.009 &    0.042/  0.023 &  I  \\
    354.9508 &   1.236 &   1.50 &   --0.72 &    2.585 &    2.585/2.611 &    2.567/2.591 &    2.599/2.625 &     0.000/--0.025 &    0.018/--0.006 &  F  \\
    355.1951 &   0.095 &   5.80 &   --0.36 &    2.773 &    2.604/2.826 &    2.561/2.781 &    2.632/2.853 &     0.169/--0.054 &    0.212/--0.008 &  I  \\
    355.1951 &   0.095 &   6.50 &   --0.36 &    2.806 &    2.795/3.012 &    2.746/2.958 &    2.823/3.037 &     0.011/--0.206 &    0.060/--0.152 &  F  \\
    360.7369 &   1.236 &   1.30 &   --0.70 &    2.587 &    2.560/2.577 &    2.546/2.563 &    2.578/2.596 &     0.028/  0.010 &    0.041/  0.024 &  I  \\
    360.7369 &   1.236 &   1.40 &   --0.70 &    2.520 &    2.519/2.542 &    2.501/2.523 &    2.533/2.556 &     0.001/--0.022 &    0.019/--0.003 &  F  \\
    367.1264 &   0.713 &   3.40 &   --0.58 &    2.555 &    2.505/2.587 &    2.493/2.573 &    2.538/2.623 &     0.051/--0.032 &    0.062/--0.017 &  I  \\
    367.1264 &   0.713 &   3.60 &   --0.58 &    2.493 &    2.513/2.616 &    2.489/2.588 &    2.542/2.648 &   --0.019/--0.123 &    0.005/--0.095 &  F  \\
    371.4777 &   0.527 &   3.00 &   --0.96 &    2.633 &    2.593/2.656 &    2.582/2.642 &    2.625/2.689 &     0.040/--0.022 &    0.051/--0.009 &  I  \\
    371.4777 &   0.527 &   3.20 &   --0.96 &    2.572 &    2.591/2.671 &    2.569/2.646 &    2.618/2.700 &   --0.019/--0.099 &    0.003/--0.073 &  F  \\
    408.5719 &   0.931 &   0.35 &   --1.84 &    2.652 &    2.637/2.640 &    2.635/2.638 &    2.664/2.668 &     0.015/  0.011 &    0.017/  0.014 &  I  \\
    408.5719 &   0.931 &   0.37 &   --1.84 &    2.582 &    2.581/2.586 &    2.571/2.575 &    2.602/2.606 &     0.001/--0.003 &    0.011/  0.007 &  F  \\
\noalign{\smallskip}
\hline
\end{tabular}
\end{table*}

\section{Analysis of a selected subsample of \ion{Zr}{ii} lines}

We particularly investigated the \ion{Zr}{ii} lines selected by \citet{ljung06}
for the solar abundance determination. 
We find this sample interesting because it contains:
two mostly clean lines ($\lambda$\, 405.0, 420.8\,nm),
a line blended on the blue side ($\lambda$\, 449.6\,nm),
two lines blended on the red side ($\lambda$\, 402.4, 425.8\,nm), and
two lines blended on both sides ($\lambda$\, 444.2, 511.2\,nm)

\subsection{The \ion{Zr}{ii} line at 402.4\,nm}
\begin{figure}
\includegraphics[width=80mm,clip=true]{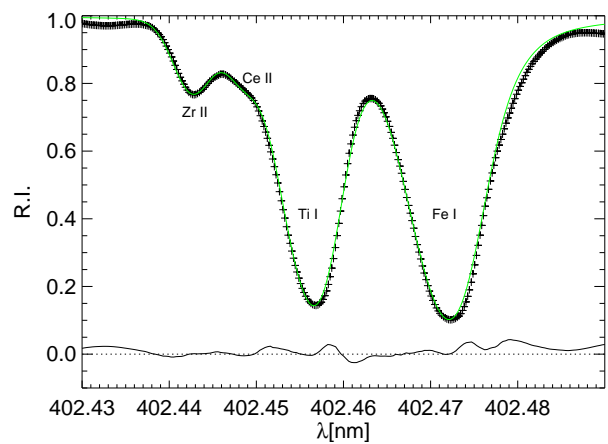}
\caption{The observed solar disc-centre spectrum of \citet{delbouille} 
in the region of the \ion{Zr}{ii} line at 402.4\,nm (black  symbols),
superimposed on the best fitting 3D synthetic profile obtained with 
method 'B' (solid green/grey).
The graph near the bottom shows the synthetic--observed residuals.
The Zr abundance derived from this fit is A(Zr)=$2.602$.
}
\label{zrii4024}
\end{figure}

This weak \ion{Zr}{ii} line is on the blue wing of a strong blend of
\ion{Ti}{i} and \ion{Fe}{ii} where the \ion{Ti}{i} line is the dominant component.
On the blue wing of this Ti-Fe blend, there is also a \ion{Ce}{ii} line.
Close by, on the red side of the \ion{Ti}{i} line, 
there is a strong line of iron blended with a much weaker \ion{Nd}{ii} line.
The zirconium abundance is determined with method 'B' 
(see Sect.\,\ref{s:methods}):
to take into account the effects of the lines present in the range
of the \ion{Zr}{ii} line, we computed a 3D profile for each of the \ion{Ce}{ii} 
line at 402.44, the \ion{Ti}{i} line at 402.45\,nm,
and the strong \ion{Fe}{i} line at 402.47\,nm, to be used in the fitting
process. The comparison of the best fitting 3D synthetic profile and the 
observed \citet{delbouille} disc-centre spectrum is shown in 
Fig.\,\ref{zrii4024}.
The Zr abundances and EWs resulting from the best fit of the four
observed spectra are listed in Table\,\ref{zr9abbo}.

The situation of the $402.4$\,nm line may be compared to that of 
the 425.8\,nm \ion{Zr}{ii} line (see Sect.\,\ref{szrii4258}),
and the $444.2$\,nm \ion{Zr}{ii} line (see Sect\,\ref{szrii4442}), 
for which we show below that the abundance is underestimated 
by about 0.03\,dex and 0.01 dex, respectively, when using the 
approximate method 'B'.

The use of {\tt splot} for the EW measurements is difficult in this case,
also with the deblending option. In fact, several lines should be introduced 
in the range, making the profile fitting unstable.

\begin{figure}
\includegraphics[width=80mm,clip=true]{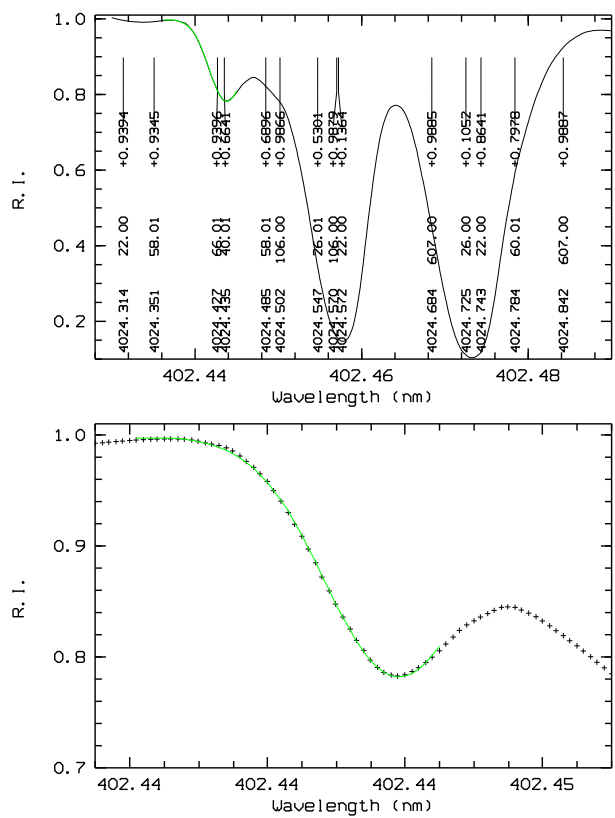}
\caption{In the top panel, the observed solar disc-centre spectrum of \citet{delbouille} 
in the region of the \ion{Zr}{ii} line at 402.4\,pm (black symbols), 
superimposed on the best fitting 3D synthetic profile (solid green/grey)
obtained with a grid computed with only zirconium.
In this fit A(Zr)=2.644. From the fit of the \citet{neckelobs} 
disc-centre profile, we get A(Zr)=2.669.
The lower panel shows a zoom in the fitting range.
}
\label{zrii4024fit}
\end{figure}

We also fitted the line profile with a grid of 3D synthetic spectra
computed with \emph{zirconium only}. We limited the fitting to the blue
wing and the line core, neglecting the red wing, which is blended
with the other lines (see Fig.\,\ref{zrii4024fit}). 
In this kind of fit, neglecting the contribution of the other
lines, the continuum placement, when considered as a free parameter, 
is problematic. When fixing the continuum, the abundance 
from the fit is in close agreement with the abundance determination 
from method 'B', in the case of the disc-centre spectra within 0.03\,dex
(cf.\  Fig.\,\ref{zrii4024fit}).
For the disc-integrated spectra we find the fit to be ambiguous. 
The abundance comes out larger than from method 'B' by about 0.1\,dex, 
but the fit is convincing only when taking the atlas of \citet{neckelobs},
although even in this case the core of the line is not well fitted.
We find a different result from the two disc-integrated and 
the two disc-centre spectra, respectively, that we cannot easily explain.
Maybe the differences are related to the fact that lines in 
flux spectra are more broadened than at disc-centre.

\subsection{The \ion{Zr}{ii} line at 405.0\,nm}
\begin{figure}
\includegraphics[width=80mm,clip=true]{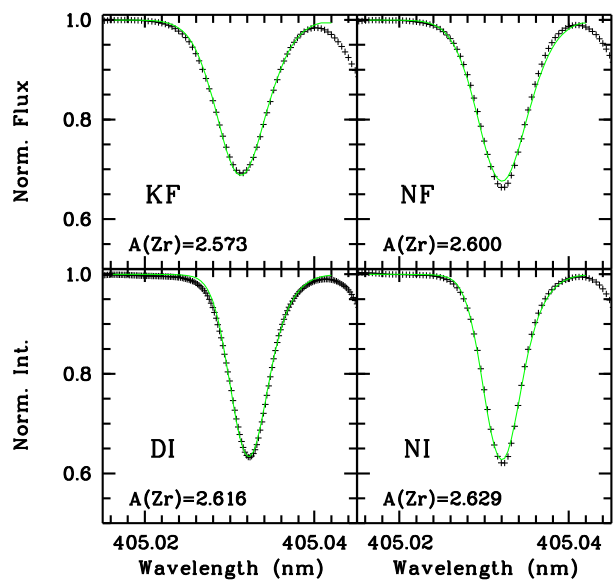}
\caption{The observed solar disc-integrated (top) and disc-centre (bottom)
spectra in the region of the \ion{Zr}{ii} line at 405.0\,pm (black symbols), 
superimposed on the best fitting 3D synthetic profile (solid green/grey)
obtained with a grid computed with only zirconium.
}
\label{zrii4050fit}
\end{figure}

\begin{figure}
\includegraphics[width=80mm,clip=true]{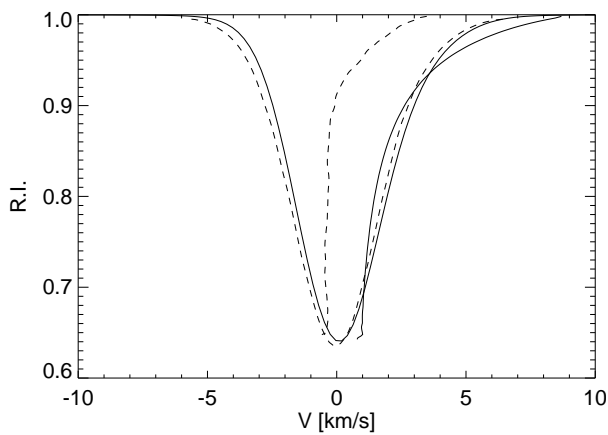}
\caption{The observed solar spectrum of \citet{delbouille}
for the the \ion{Zr}{ii} line at 405.0\,pm (dashed black),
together with its line bisector (plotted on a ten-times expanded abscissa),
superimposed on the 3D synthetic profile (solid black)
with its line bisector.
}
\label{zrii4050bisector}
\end{figure}

This line seems to be unblended. A weaker line is on the red 
side at about $\Delta \lambda$ 0.016\,nm from the centre of the \ion{Zr}{ii}
line, and a stronger \ion{Fe}{i} line is also on the red side at about 
$\Delta \lambda$ 0.035\,nm.
Looking at the spectra one could conclude that the line is clean and that
the abundance can be deduced from line profile fitting with a synthetic
grid computed with only Zr and/or by easily measuring the EW.
When we fit the line profile with a grid that takes 
into account only zirconium, the result is very satisfactory for all observed 
spectra, except perhaps for the disc-integrated spectrum of \citet{neckelobs}
(see `NF', Fig.\,\ref{zrii4050fit}). The results are in close agreement with 
the abundance determinations from the EWs measured with {\tt splot}.
We also compared the theoretical line bisector from the 3D synthetic line
profile to the observed disc-centre spectrum of \citet{delbouille}, and we 
find an excellent agreement (see Fig.\,\ref{zrii4050bisector}).
The shift in wavelength between observed and synthetic line bisector
is mainly due to the fact that the spectrum of \citet{delbouille} is 
not corrected for gravitational redshift.

\subsection{The \ion{Zr}{ii} line at 420.8\,nm}
\begin{figure}
\includegraphics[width=80mm,clip=true]{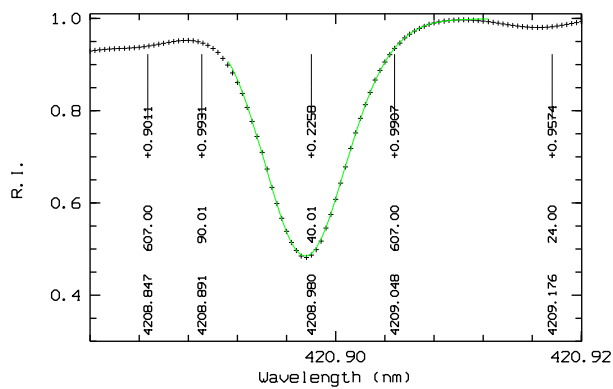}
\caption{The observed solar disc-integrated spectrum of \citet{kuruczflux}
(black symbols) in the region of the \ion{Zr}{ii} line
at 420.8\,pm, superimposed on the best fitting 
profile (solid green/grey) derived from a grid of 3D synthetic 
spectra computed with only zirconium. 
The fit corresponds to A(Zr)=2.58. 
}
\label{zrii4208}
\end{figure}

This line is strong and sensitive to the adopted damping constants.
The red side is more or less clean, while on the blue side there
is a weak line, and close by another weak line.
The line profile fitting is done with method 'A', using Zr only, 
and restricting the fit to the wavelength range $420.891$ \ldots 
$420.913$\,nm in order to avoid the weak blends in the far blue
wing of this \ion{Zr}{ii} line. The continuum level is fixed in the 
fitting procedure.
The fitting of the disc-integrated line profile in the \citet{kuruczflux} 
spectral atlas is shown in Fig.\,\ref{zrii4208}.
The abundance derived from line profile fitting and from the EW measured
with {\tt splot} are in very good agreement (within 0.007\,dex) for the 
observed spectra of \citet{delbouille} and \citet{kuruczflux}. For
\citet{neckelobs}, the abundance from line profile fitting is 
lower with respect to the abundance derived from the EW 
measurements by 0.04\,dex and and 0.01\,dex, for the disc-integrated 
and disc-centre spectrum, respectively.

\subsection{The \ion{Zr}{ii} line at 425.8\,nm} \label{szrii4258}
\begin{figure}
\includegraphics[width=80mm,clip=true]{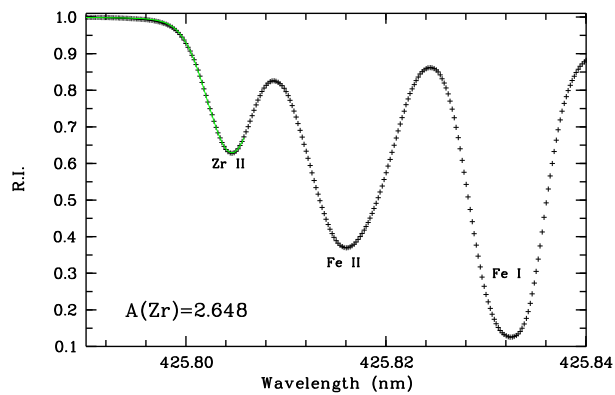}
\caption{The observed solar disc-centre spectrum of \citet{delbouille} 
in the region of the \ion{Zr}{ii} line at 425.8\,pm (black symbols) 
is superimposed to the 3D best fit (solid green/grey), obtained
with a grid of 3D synthetic profiles computed with only Zr.
}
\label{zrii4258}
\end{figure}

This \ion{Zr}{ii} line is on the blue wing of a much stronger \ion{Fe}{ii}
line that has close by, on the red side, a very strong \ion{Fe}{i} line,
blended with a weaker \ion{Fe}{ii} line (see Fig.\,\ref{zrii4258}).
We tried different fitting procedures. 

First, we fitted the \ion{Zr}{ii} 
line profile with a grid of 3D synthetic profiles computed with only Zr, 
restricting the fitting to the blue wing of the Zr line. For the disc-centre 
spectrum of \citet{delbouille} we obtain A(Zr)=$2.65\pm 0.01$. The fit looks 
quite satisfactory (see Fig.\,\ref{zrii4258}), but it is not clear how much 
the strong lines close by affect the \ion{Zr}{ii} line, and hence by how 
much A(Zr) is overestimated.

Second, we applied method 'B', taking into account the contribution of
the neighbouring iron lines as a linear combination of scaled 3D
synthetic profiles, and obtain A(Zr)=$2.61\pm 0.01$.

\begin{figure}
\includegraphics[width=80mm,clip=true]{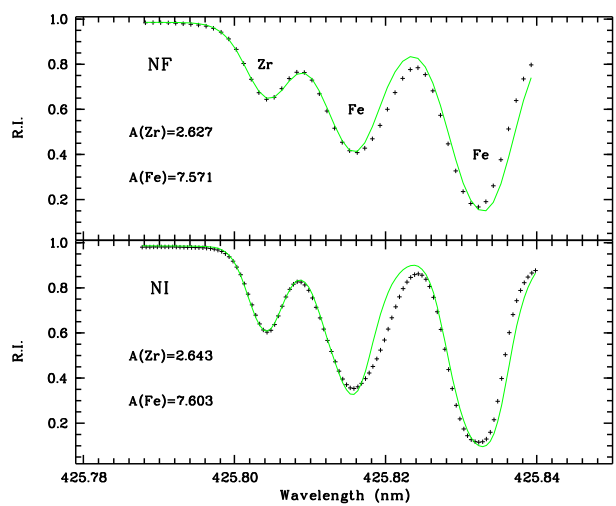}
\caption{The observed \citet{neckelobs} spectrum
in the region of the \ion{Zr}{ii} line at 425.8\,pm (black symbols) 
is superimposed to the best fitting 3D synthetic spectrum (solid green/grey)
obtained with method 'A'. Upper panel: disc-integrated spectrum,
lower panel: disc-centre spectrum.
}
\label{zr4258_fit3d}
\end{figure}

Finally, we also fitted the complete range with method 'A', properly
taking into account the three iron lines, with A(Zr) and A(Fe) as free
fitting parameters.  We obtain a Zr abundance of $2.643\pm 0.001$ and
$2.627\pm 0.000$ from disc-centre and disc-integrated spectra,
respectively.  Even though the iron lines are not very well reproduced
(see Fig.\,\ref{zr4258_fit3d}), the \ion{Zr}{ii} line is fitted very
well and is not affected by the minor deficiencies of the iron lines.
The equivalent widths given in Table\,\ref{zr9abbo} are derived from
the abundances obtained with this fitting method.

\subsection{The \ion{Zr}{ii} line at 444.2\,nm} \label{szrii4442}
This line is surrounded by two stronger features, an iron line on the
blue side at 444.2831\,nm and a blend of three lines of iron on the
red side.  Of these three \ion{Fe}{i} lines on the red side, the one
at 444.3194\,nm is much stronger than the other two
components. 

Clearly, the \ion{Zr}{ii} line is affected by the 
two \ion{Fe}{i} features on both sides, and cannot be treated in isolation.
Since the two iron features strongly influence the shape of the \ion{Zr}{ii} 
line, the Zr abundance cannot be derived from this line by fitting 
with a grid of 3D synthetic profiles computed with zirconium only. 

\begin{figure}
\includegraphics[width=80mm,clip=true]{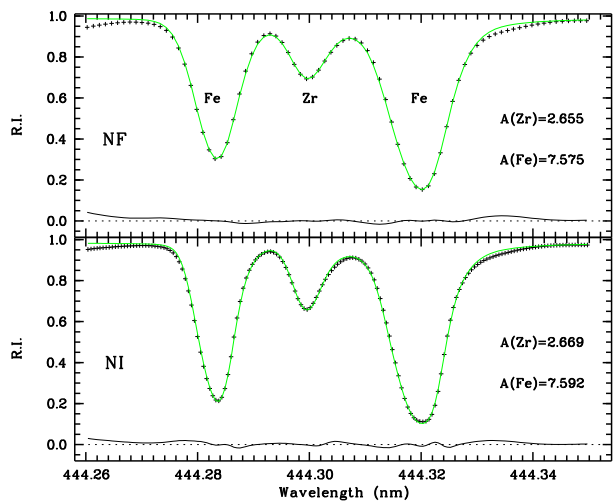}
\caption{The observed solar disc-centre spectrum of \citet{neckelobs} 
in the region of the \ion{Zr}{ii} line at 444.2\,pm (black symbols)
with the 3D synthetic best-fit-profile obtained with method 'A' 
superimposed (solid green/grey). 
Upper panel: disc-integrated spectrum, lower panel: disc-centre spectrum.
The residual is below each plot.
}
\label{zr4442_fit3d}
\end{figure}

The most reliable Zr abundance is obtained from the full fitting, method 'A':
we fitted the observed Fe--Zr--Fe blend with A(Zr) and A(Fe)
as free parameters. The Zr abundance derived in this way is 
$2.673\pm 0.004$ and $2.650\pm 0.005$ for the disc-centre and 
disc-integrated spectra, respectively (cf.\ Fig.\,\ref{zr4442_fit3d}).
The equivalent widths given in Table\,\ref{zr9abbo} are derived from 
the abundances of this fitting method.

For completeness, we also derived the Zr abundance with methods 'B'
and 'C'.

 For method 'B', the profiles of the four iron blending features
are represented by scaled 3D synthetic Fe line profiles, which are
then added to the profile of the \ion{Zr}{ii} line in order to
obtain an approximate profile of the complete blend (ignoring
saturation effects). With this approach, the best fits are obtained
with Zr abundances of A(Zr)=$2.67\pm 0.01$ and A(Zr)=$2.64\pm 0.01$
for disc-centre and disc-integrated case, respectively.
The fit of the observed disc-integrated spectrum is somewhat better
that in the case of the disc-centre observations.

In method 'C', A(Zr) is found from the equivalent width of the
\ion{Zr}{ii} line, as obtained from {\tt splot} with the deblending
option.  Representing the Fe feature on the red side of the
\ion{Zr}{ii} line either by a single or by three Voigt profiles turned
out to have no influence on the resulting zirconium abundance.
We find A(Zr) derived in this way to be very close to the result of 
method 'B' in the case of the disc-centre observed spectra.
In the disc-integrated case, A(Zr) is about $0.057$\,dex smaller
than what is obtained with method 'B'.

\subsection{The \ion{Zr}{ii} line at 449.6\,nm} \label{szrii4496}
This \ion{Zr}{ii} line (EW $\approx 3$\,pm) is contaminated by a 
much stronger \ion{Cr}{i} line (EW $\approx 9$\,pm).
Certainly, saturation effects cannot be ignored in this case.

\begin{figure}
\includegraphics[width=80mm,clip=true]{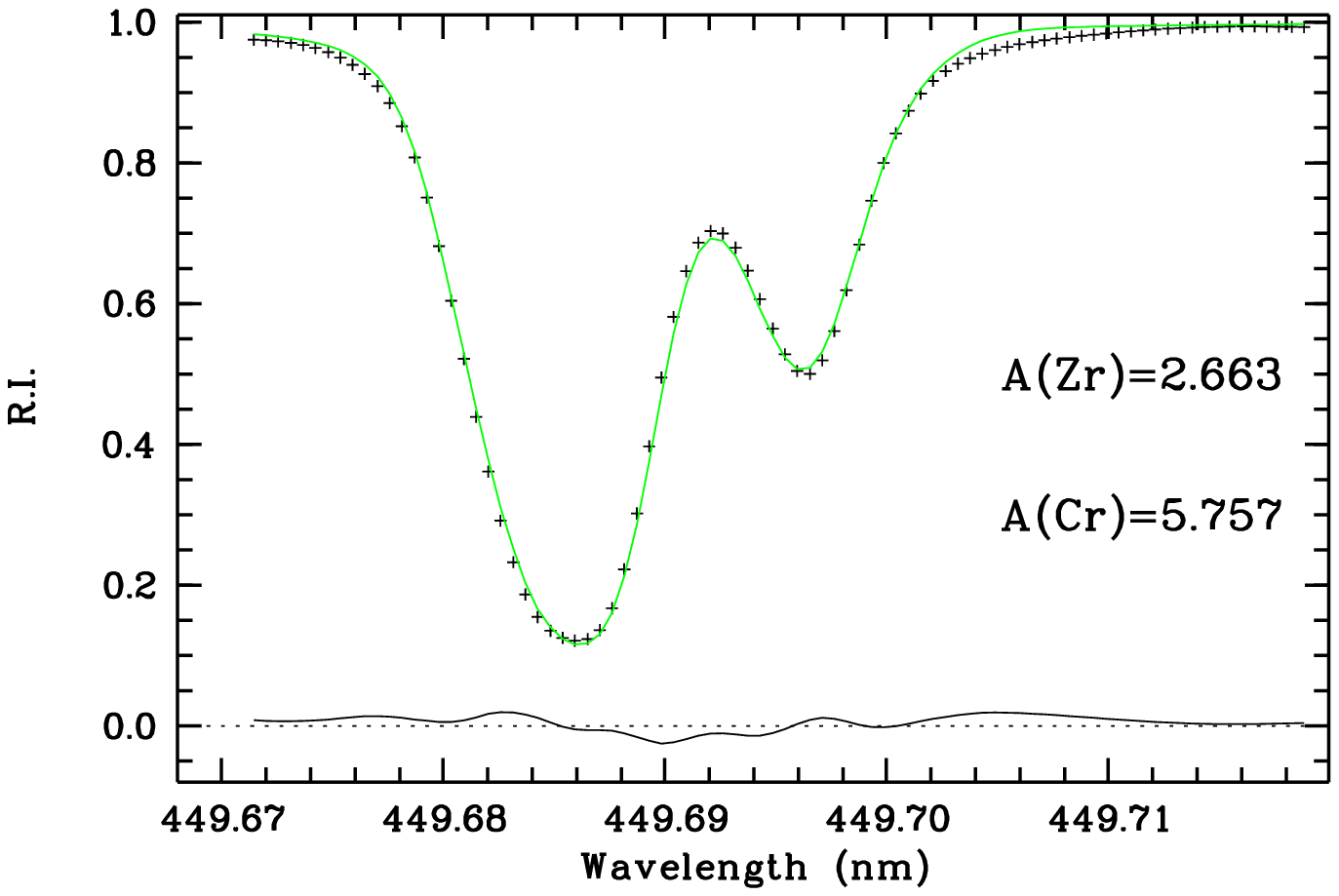}
\caption{The observed solar disc-centre spectrum of \citet{neckelobs} 
in the region of the \ion{Zr}{ii} line at $449.6$\,nm (black symbols) 
with the best fitting 3D synthetic profile obtained from method 'A' 
superimposed (solid green/grey). 
The graph in the lower part shows the synthetic $-$ observed residuals.
}
\label{zr4496_fit3d}
\end{figure}

Based on the full 3D line profile fitting of the complete blend using
method 'A', with A(Cr) and A(Zr) as free fitting parameters, we obtain 
a Zr abundance of A(r)=$2.660\pm 0.003$ for disc-centre (cf.\
Fig.\,\ref{zr4496_fit3d}), and A(Zr)=$2.642\pm 0.005$ for integrated
disc. These Zr abundances correspond to equivalent widths of 3.11 and
3.41\,pm respectively (see Table\,\ref{zr9abbo}).

\begin{figure}
\includegraphics[width=80mm,clip=true]{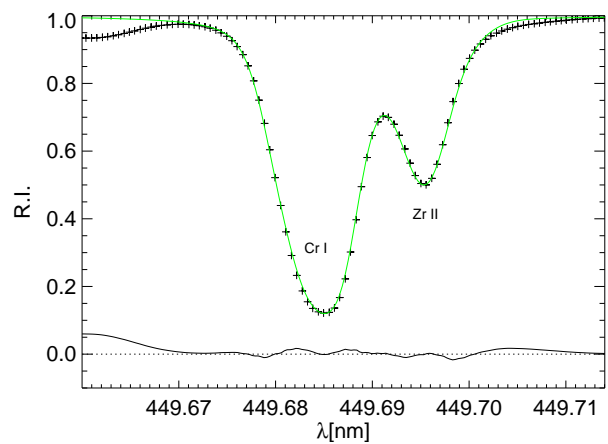}
\caption{The observed solar disc-centre spectrum of \citet{neckelobs} 
in the region of the \ion{Zr}{ii} line at $449.6$\,nm (black symbols) 
with the best fitting 3D synthetic profile obtained from method 'B' 
with A(Zr)=2.63 superimposed (solid green/grey). 
The graph in the lower part shows the synthetic $-$ observed residuals.
}
\label{zrii4496}
\end{figure}

The fit resulting from method 'B', scaling the Cr profile and
interpolating in the grid of 3D synthetic Zr profiles to achieve the
best agreement between synthetic and observed profile, is shown in
Fig.\,\ref{zrii4496} for the disc-centre case. It is of similar
quality as the one obtained from the full fitting procedure with
method 'A' (Fig.\,\ref{zr4496_fit3d}). Method 'B' gives A(Zr)=$2.63$
and $2.60$ for disc-centre and integrated disc, respectively.
As expected, the approximate method 'B' tends to underestimate the 
Zr abundance, in this example by $0.027$ and $0.042$\,dex
for disc-centre and integrated disc, respectively.
The average equivalent width we derive with this method for the 
observed disc-centre and disc-integrated spectra are 
EW=$3.01$ and $3.25$\,pm, respectively, $3.2$ and $4.7$\% lower 
compared to method 'A'.

When we apply method 'C' using {\tt splot} with the deblending option,
we cannot reproduce the Cr line profile with a single Voigt profile, but
we can formally have a good reproduction of the profile with two Voigt
profiles, but this option cannot take into account properly the 
wings of the \ion{Cr}{ii} line.

\subsection{The \ion{Zr}{ii} line at 511.2\,nm}
\begin{figure}
\includegraphics[width=80mm,clip=true]{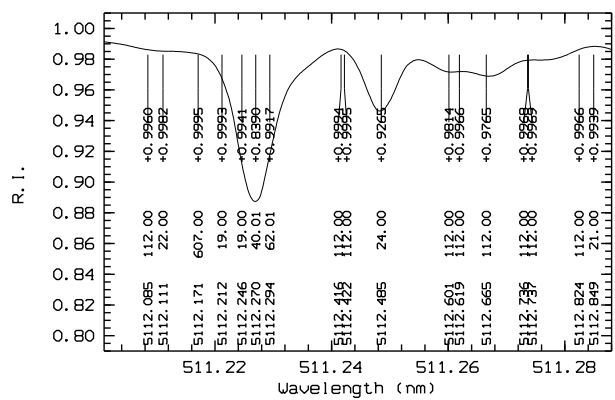}
\caption{The observed solar disc-centre spectrum of \citet{delbouille} 
in the region of the \ion{Zr}{ii} line at 511.2\,nm (black solid) with 
the identifications of the lines in the range according the line list of 
Kurucz.
}
\label{zrii5112}
\end{figure}

For this line, the placement of the continuum is difficult, 
but the uncertainty induces an error in the  
EW measurement of less than 
4\% and hence a difference in the Zr abundance of at most 0.02\,dex.
The line shape suggests that it is blended on both wings,
and in fact three weak blend lines are expected according to the Kurucz line 
list \footnote{\url{http://kurucz.harvard.edu/linelists.html}};
two \ion{K}{i} lines at 511.2212 and 511.2246\,nm, and a \ion{Sm}{ii} 
line at 511.2294\,nm. In addition, several weak molecular lines are expected 
in the range (see Fig.\,\ref{zrii5112}).

The average EWs derived with method 'B' for disc-centre 
and integrated disc, respectively, are 8.4 and 9.2\,pm, corresponding to
zirconium abundances of A(Zr)=$2.69$ and $2.67$.
When measuring the EW with {\tt splot} (method 'C'), we find similar results, 
but the uncertainty in the measurements is larger, giving an indetermination 
in the abundance determination of about 0.06\,dex.

\section{Conclusions}

We analysed the zirconium abundance in the solar photosphere,
investigating a selected sample of lines of both \ion{Zr}{i} and 
\ion{Zr}{ii}. 
The \ion{Zr}{i} lines are weak and heavily blended so 
that only four of them are acceptable for abundance work.
However, the present analysis of the zirconium abundance 
relies primarily on 15 lines of \ion{Zr}{ii} that we found suitable for 
this purpose. 

We have applied three different fitting strategies to derive the
abundance from Zr lines that are blended by lines of other elements.
In the case that all components making up the blend are weak, the
different methods give consistent abundances. If, however, stronger
lines are involved (for an example see Sect.\,\ref{szrii4496}), the
methods that ignore saturation effects may severely underestimate the
Zr abundance by up to $0.05$\,dex, even though the result of the
fitting may appear pleasing to the eye, and the reduced $\chi^2$ may
be close to one.

We find a good agreement between A(Zr) derived from the \ion{Zr}{i} and the 
\ion{Zr}{ii} lines, but, due to the scarcity of \ion{Zr}{i} lines, we
consider this result as fortuitous.
The abundance from disc-centre spectra is systematically higher than 
the one from disc-integrated observations. A similar result is also found 
for other heavy elements like Fe, Th, Hf, both with 3D and 1D model 
atmospheres. This may indicate a problem with the thermal structure of the 
models, rather than a physical abundance gradient in the solar atmosphere.
Further investigations are necessary to find an explanation for this small 
discrepancy.

Our recommended solar zirconium abundance 
is based on the 3D result for 15 \ion{Zr}{ii} lines, and is
A(Zr)$=2.62\pm 0.06$, where the uncertainty is the line-to-line 
scatter of the selected sample of \ion{Zr}{ii} lines. 
This value is at the upper end of the solar zirconium abundances
found in the literature. Still, within the mutual error bars, 
this result is in good agreement with the meteoritic zirconium
abundance of A(Zr)=$2.57\pm 0.04$ \citep{lodders09}.


\acknowledgements
We acknowledge financial support from 
``Programme National de Physique Stellaire'' (PNPS) and
``Programme Nationale de Cosmologie et Galaxies'' (PNCG) 
of CNRS/INSU, France.
We thank the referee, K. Lodders, for helpful comments on the
relation between cosmochemical and astronomical abundance
scale.




\end{document}